\documentclass[aps,showkeywords,groupedaddress,nofootinbib]{revtex4}
\usepackage{graphicx}
\usepackage{color}
\usepackage{amsmath}
\usepackage{amsfonts}
\usepackage[toc]{appendix}


\begin{document}

\title{Effect of Overdispersion of Lethal Lesions on Cell Survival Curves}

\author{M. Loan$^{a}$\footnote{Corresponding author} and A. Bhat$^{b}$}
\affiliation{$^{a}$ANUC, Australian National University, Canberra, 2600, Australia\\
$^{b}$Department of Oncology, East Tennessee State University, Tennessee, 37614, USA}

\date{\today}

\begin{abstract}
\noindent {\bf Purpose:} The linear-quadratic (LQ) model is one of the most commonly used tools for predicting radiotherapy outcomes and
has been used extensively to analyse dose responses to ionizing radiation both in vitro and in vivo. However, there remain questions about the wider applicability of the LQ model in terms of its representative nature of some deeper mechanistic behaviour. In particular, empirical evidence suggests that the LQ model tends to underestimate cell survival at low doses and overestimate cell killing at high doses. It is believed to be driven from the usual LQ model assumption that radiogenic lesions are Poisson distributed. In this context, we explore the effects of over dispersed DNA lesion distribution on the shapes of cell surviving curves of mammalian cells exposed to hadrons at various doses.\\
\noindent {\bf Methods:} To provide a theoretical framework in resolving discrepancies between experimental data and LQ model predictions, we employ a non-Poisson distribution of lethal lesions together with non-homologous end-joining (NHEJ) pathway of double-strand break (DSB) repair. A negative binomial (NB)
distribution is used to study the effect of the overdispersion on the shapes and possible reduction of dose-response curvature at high doses. The error distribution is customized to include an adjustable parameter so that the overdispersion parameter of NB is not constant but depends on the mean of the distribution. The trends in predicted cell survival responses are compared with the experimental data in low and high dose regions at various LET values for proton, helium, and carbon ions.\\
\noindent{\bf Results and Conclusions:}
The cell survival responses calculated by the present method reveal straightening of survival curves at high doses. This suggests that the overdispersion causes the cell survival dose-response to approximate log-linear behaviour at high doses. Comparison of the cell survival predictions with the Particle Irradiation Data Ensemble (PIDE) shows that the NB model provides better fits to the experimental data following low and intermediate doses. Whereas the model predictions are not validated at tiny and very high doses, nonetheless, the presented approach provides insight into underlying microscopic mechanisms which may help to improve the radiobiological responses along the dose-response curves and resolve discrepancies between experimental data and current cell survival models.\\

\vspace{0.20cm}

\end{abstract}

\maketitle

\newpage
\section{Introduction}
It has long been established that the linear-quadratic (LQ) model is a preferred mechanism for quantitative predictions of the description of radiation-induced cell death and chromosomal aberration dose responses using sufficiently few adjustable parameters. Underlying the application of the LQ model to fractionation effects is pairwise misrepair of primary lesions, which are resolved either through restitution or binary misrepair. It is assumed that the frequency of chromosomal aberrations is a linear-quadratic function of dose because the aberrations are consequences of the interaction of two separate DNA breaks. At low doses per fraction, both breaks may be caused by the same ionization event; the probability of exchange aberration is proportional to the dose. At the higher doses, the two breaks are more likely to be caused by multiple hits, the probability of exchange aberration is proportional to the square of the dose. The model has also proved to be a good approximation to a wide range of damage-kinetic models which describe the kinetics of DNA double-strand breaks
\cite{Thames1985,Preston1990,Gerweck1994,Sachs1997,Stewart2001,Guerrero2002,Brown2003}. Whereas these models have different formulations for the dose dependence of the mean lethal lesion yield, the common feature of explaining clonogenic cell survival is the Poisson distribution of the mean number of lethal lesions per cell. Such assumptions are understandable, given that these models evolved to explain clonogenic cell survival, an endpoint
for which the number of lesions in individual cells cannot be quantified. Overall, these mechanisms produce a linear-quadratic-linear dose-response
relationship at low LET \cite{Curtis1986,Rossi1988,Brenner1990,Obaturov1993,Tobias1995,Hawkins1996,Radivoy1998,Brenner1998,Carlone2005}.\\

\noindent However, despite the widespread usage of the LQ model, there remain questions about its wider applicability \cite{Sachs1998,Zaider1998,Kirk2009,McMohan2019}. For example, there is growing evidence that indicates that the LQ model exhibits a departure from of its linear quadratic relation at very low and very high doses. It has been observed that in the phenomenon of low-dose hypersensitivity, some cells show dramatic sensitivity to low doses $< 0.5$ Gy, before returning to an LQ-like response at higher doses \cite{Joiner2001}. Studies have shown that the dose-response curve begins to straighten at higher doses, leading some investigators to propose that a linear-quadratic-linear dose-response model is more appropriate \cite{Astrahan2008}. The discrepancy between the measured cell survival data and the LQ model predictions at a higher dose is explained by the heterogeneity of sensitivity to radiation among the cells of the irradiated population \cite{Hawkins2017b,Ward1988,Goodhead1994,Heilmann1995}. Postulating that increasing LET causes non-random clustering of lethal lesions in some cells to deviate from the Poisson distribution, Hawkins reported a detailed study on the effect of deviation from the Poisson distribution of lethal lesions on the dependence of RBE in the limit of zero dose on the LET greater than that of linear range \cite{Hawkins2003,Hawkins2017,Loan2020}. Using a Compound Poisson distribution, Nowak \emph{et al.,} modelled the clustering of breakage events as the process leading to non-exponential “spacing” between subsequent events, to relate parameters of such distributions to relevant quantities describing the number of induced DSBs \cite{Nowak2000,Virsik1981,Goodwin1994}. More recently, Shuryak \cite{Shuryak2017,Shuryak2020} reported the evidence of overdispersion by comparing the fits to the fibroblasts data from a Poisson distribution and the Negative Binomial distribution. The study reported that the Poisson distribution underestimated the “upper tail” of the observed distribution.\\

\noindent Using a non-Poisson approach as a more flexible alternative that allows accommodating a variety of mechanisms for overdispersion, we employ a customized negative binomial (NB) distribution to calculate the survival probabilities of V79 cell lines radiated by hadrons and light ions. For higher LET radiations, the individual particle tracks can infer multiple instances of lethal damage, and the ratio of the variance of the distribution to the mean of the distribution exceeds unity. The use of such a non-Poisson distribution implicitly determines the density and spatial distribution of DSB within and among cells. That is, DSB formed along high LET particle tracks tends to be in closer spatial proximity than DSB formed by different particle tracks. The rest of the paper is organised as follows: In Sec.II, we develop an NB error model for cell survival fraction in terms of modified radiosensitive parameters. The DSB repair kinetics is described using the non-homologous end-joining pathway. Biologically relevant quantities, such as radiosensitive parameters, are further derived and used to obtain the fit parameters of the model. The model is tested against PIDE and other observed data in Sec.III. The results of these comparisons provide useful information on how changing the error distribution for radiation-induced lethal lesions from Poisson to NB, without changing the LQ dose dependence for the mean, alters the performance of the LQ model in describing survival curve predictions at high doses and low doses/dose rates.

\section{Model Calculation of Survival Fraction}

\subsection{Radiation Induced Average DSB Yield}

Following the LQ model, the mean number of lethal lesions per cell, $\lambda$, is described by
\begin{equation}
\lambda= -\alpha D - \beta D^{2},
\label{eqn1}
\end{equation}
where $\alpha$ and $\beta$ are the model parameters describing the radiosensitivity of cell and $D$ is the radiation dose. Assuming the lethal lesions are Poisson distributed from cell to
cell, the probability of $k$ lethal lesions in a cell is given by
\begin{equation}
P_{Pois.}(k) = \frac{\lambda^{k}e^{-\lambda}}{k!}.
\label{eqn2}
\end{equation}
and cell surviving probability is given by
\begin{equation}
SF_{Pois.} = P_{Pois.}(k=0)= e^{-\lambda}.
\label{eqn3}
\end{equation}
It has been well established that the effective plot of Eq. (\ref{eqn3}), on a log
scale, gives what is referred to as a ''shouldered'' dose-response
curve. The initial region of the curve is dominated by the linear
term at low doses, followed by increasing curvature as the
quadratic term becomes more significant. The degree of curvature
is commonly expressed in terms of the $\alpha / \beta$ ratio and corresponds to the dose at
which the linear and quadratic contributions are equal. To incorporate the effects of the overdispersed lethal chromosomal lesions
among cells, we employ a customized negative binomial
(NB) distribution. The probability of observing $k$ lethal lesions in a cell in such an error distribution is described by
\begin{equation}
P_{NB}(k) = \frac{\Gamma (k+1/\omega)}{\Gamma (1/\omega)\times k!}
\bigg(\frac{1}{1+\omega\lambda }\bigg)^{1/\omega} \times
\bigg(\frac{1}{1+1/\omega\lambda}\bigg)^{k}.
\label{eqn4}
\end{equation}
where $\omega= r(1-e^{-\lambda})$ is the modified dispersion parameter relative to the Poisson distribution and describes the dependence of bare dispersion parameter $r$ on the mean $\lambda$ that results from the possibility that at low $\lambda$ overdispersion can be negligibly small but is expected to increase to a certain limit at higher values of $\lambda$. It is expected that small values of $r$ (close to zero) generate Poisson distribution-like behaviours, whereas larger $r$ values produce distributions with larger tails \cite{Shuryak2017}. The NB error modelled surviving fraction becomes
\begin{equation}
SF_{NB} =P_{NB}(k=0)= (1+\omega\lambda)^{-1/\omega}.
\label{eqn5}
\end{equation}

\noindent Following \cite{Wang2018}, we first calculate the average number of primary particles that cause
DSB, $n_{p}$, and the average number of DSBs yielded by each
primary particle that causes DSB, $\lambda_{p}$. The DSB yield per
cell per primary particle is given by $\lambda = N/n$, where
$N=YD$ is the average number of radiation induced DSBs per cell
and $n$ is the number of the particles passing through the cell
nucleus
\begin{equation}
n = \frac{\pi R^{2}D\rho}{0.160\times LET},
\label{eqn6}
\end{equation}
where $R$ is the cell nucleus radius ($\mu$m) and $\rho$ is density (g/$\mbox{cm}^{3}$). Assuming that the number of DSBs yielded by a primary
particle is NB distributed, the probability of a primary particle passing
through a nucleus without causing any DSB is given by
\begin{equation}
P_{NB}(k=0)= (1+\omega \lambda)^{-1/\omega}.
\label{eqn7}
\end{equation}
The average number of primary particles that cause DSB, $n_{p}$, and the
average number of DSBs yield per primary particle that causes DSB, $\lambda_{p}$,
are then given by
\begin{eqnarray}
n_{p} & = & n[1-(1+\omega \lambda )^{-1/\omega}], \label{eqn8a}\\
\lambda_{p} & = & \frac{\lambda}{[1-(1+\omega \lambda )^{-1/\omega}]}.
\label{eqn8}
\end{eqnarray}

\subsection{Cell Survival Curve}
Double strand breaks in DNA form as a result of exposure to exogenous agents such as radiation and certain chemicals, as well as through endogenous processes, including DNA replication and repair. Three DSB repair processes, the non-homologous end-joining (NHEJ) pathway, the homologous recombination repair (HRR) pathway, and the Mismatch End Joining (MMEJ) are commonly used in mammalian cells with NHEJ likely playing the largest role in DSB repair \cite{Karge2017,Burma2006,Malu2012}. NHEJ mediates the direct relegation of the broken DNA molecule \cite{Weterings2008} and has the potential to bind together any type of DNA ends and does not require a homologous template for repair of the DNA lesion. Unlike HRR, the non-homologous end-joining (NHEJ) pathway is not restricted to a certain phase of the cell cycle. Recent studies \cite{Baiocco2016, Friedland2017} indicate that the relationship between radiation dose and the number of DNA fragments shorter than 30 bp induced by radiation is considered to be linear. This indicates that, other than the interaction among DSBs induced by different primary particles, both the clustered DNA damage effect (two or more DSBs within a typical distance of 25 bp) and the overkill effect (two or more DSBs within 10 bp) of high LET radiation depend mainly on the DSB distribution on the track of the primary particles. Therefore, both effects depend on the average number of DSBs yield by each primary particle that caused DSB, $\lambda_{p}$. The average number of lethal events, $N_{p}$, can be modelled by
\begin{equation}
N_{avg} =N\times (1- P_{correct}),
\label{eqn10}
\end{equation}
where $P_{correct}$ is the total probability of a DSB being correctly rejoined.
\noindent For any break in a particular condition, the true distribution of rejoining rates is non-trivial and lacks a simple form for its overall distribution. Following \cite{McMahon2016}, the repair behaviour is determined by approximating the recombination function to a step function where the break in question will have the same recombination rate for all potential breaks occurring at distance $d_{max.}$ within a shell of radius R. This simplifies the probability of correct repair for a given break to $1/(1+k )$, where $k$ is the number of breaks within the distance $d_{max}$ \cite{McMahon2016}. \\

\noindent Following the general mechanism of NHEJ, each DSB may be (i) rejoined with the other end from the same DSB, (ii) joined with a DSB end from a different DSB induced by the same primary particle, (iii) joined with a DSB end from a DSB induced by a different primary particle, and (iv) left without being joined with any DSB end. Each pathway has an associated fidelity which indicates the probability with which the pathway will correctly repair a given DSB. The average probability that the DSB is correctly joined with the other end from the same DSB is assumed to be $\mu_{x}$, which quantitatively describes the fidelity of the NHEJ pathway \cite{Burma2006,Wang2018}.

\noindent Assuming that the breaks within the spherical radius are NB distributed, the expectation value that a randomly chosen break will rejoin correctly (a DSB
end do not be joined with a DSB end from a DSB induced by a different primary particle) is given by (see the Appendix)
\begin{equation}
P_{1} = \frac{1}{\lambda_{int}(1- \omega)}\left[1- (1+
\omega\lambda_{int})^{1-1/\omega}\right], \label{eqn7}
\end{equation}
where $\lambda_{int}=\eta (\lambda_{p})n_{p}$ is the average probability of a DSB end being joined with a DSB end from a DSB induced by a different
primary particle and is proportional to the average number of primary particles which caused DSBs, $n_{p}$. The relationship between $\eta (\lambda_{p})$ and $\lambda_{p}$ is assumed as \cite{Wang2018}:
\begin{eqnarray}
\left\{ \begin{array}{cccccc}
\eta (\lambda_{p}) & = & \eta_{\lambda_{p}\rightarrow \infty} - \frac{\eta_{\lambda_{p}\rightarrow\infty} - \eta_{\lambda_{p}\rightarrow 1}}{\lambda_{p}}\nonumber\\
\lim_{\eta_{\lambda_{p}\rightarrow 1}} \eta (\lambda_{p}) &=& \eta_{\lambda_{p}\rightarrow 1}\nonumber \\
\lim_{\eta_{\lambda_{p}\rightarrow \infty}} \eta (\lambda_{p}) & = & \eta_{\lambda_{p}\rightarrow \infty}.
\end{array} \right.
\label{eqn11}
\end{eqnarray}

\noindent Also, assuming that a primary particle generates DSBs randomly on its track, the probability that a DSB end not be joined with a
DSB end from a different DSB induced by the same primary particle is given by
\begin{equation}
P_{2} = \frac{1}{\lambda_{track}(1-\omega)}\left[1- (1+
\omega\lambda_{track})^{1-1/\omega}\right], \label{eqn12}
\end{equation}
where $\lambda_{track}=\phi\lambda_{p}$ is the average probability of
a DSB end being joined with a DSB end from a DSB induced by the
same primary particle. Therefore, the total probability of a DSB
being correctly repaired is given by
\begin{eqnarray}
P_{correct} & = &
\mu_{x}P_{1}P_{2}\nonumber\\
& = & \mu_{x}\left[\frac{1}{\lambda_{int}(1-\omega)}\bigg(1- (1+\omega
\lambda_{int})^{1-1/\omega }\bigg)\right]\times
\left[\frac{1}{\lambda_{track}(1-\omega)}\bigg(1-
(1+\omega \lambda_{track})^{1-1/\omega}\bigg)\right].\nonumber\\
\label{eqn13}
\end{eqnarray}
The first term in the square brackets quantitatively describes the interaction of
DSBs induced by different primary particles and the second term describes the effect of clustered DNA damage.\\

\noindent Finally, considering the overkill effect, similar with the clustered DNA damage effect from Eq.(\ref{eqn12}), the probability of a DSB having contributed to cell death is given by:
\begin{equation}
P_{3} = \frac{1}{\lambda_{contb.}(1-\omega)}\left[1- (1+
\omega\mu_{contb.})^{1-1/\omega}\right],
\label{eqn14}
\end{equation}
where $\lambda_{contb.}= \xi\lambda_{p}$.

\noindent Using the above repairing probabilities, Eq. (\ref{eqn10}) becomes
\begin{eqnarray}
N_{avg} &= & \frac{N}{\lambda_{contb.}(1-\omega)}\left[1- (1+
\omega\lambda_{contb.})^{1-1/\omega}\right]\times
\left[1-
\lambda_{x}\left\{\frac{1}{\lambda_{int}(1-\omega)}\bigg(1- (1+\omega
\lambda_{int})^{1-1/\omega}\bigg)\right\} \right. \nonumber\\
& & \times \left. \left\{\frac{1}{\lambda_{track}(1-\omega)}\bigg(1- (1+\omega
\lambda_{track})^{1-1/\omega}\bigg)\right\}\right],
\label{eqn15}
\end{eqnarray}
and the cell survival takes the form:
\begin{equation}
S_{NB} = \bigg(1+ \omega N_{avg}\bigg)^{-1/\omega}.
= \bigg(1+\omega(\alpha_{NB} D+\beta_{NB} D^{2})\bigg)^{-1/\omega},
\label{eqn16}
\end{equation}
where
\begin{eqnarray}
\alpha_{NB} & = & \frac{Y}{\xi\lambda_{p}(1-\omega)}\bigg(1-(1+\omega\xi\lambda_{p})^{1-1/\omega}\bigg)
\left[1 - \mu_{x}\frac{1}{\phi\lambda_{p}(1-\omega)}\bigg(1-(1+\omega\phi\lambda_{p})^{1-1/\omega}\bigg)\right]\label{eqn17a}\\
\beta_{NB} & = & \frac{1}{2}\frac{\eta
(\lambda_{p})}{\lambda_{p}}\frac{\mu_{x} Y^{2}}{\xi\lambda_{p}(1-\omega)}\bigg(1-(1+\omega\xi\lambda_{p})^{1-1/\omega}\bigg)
\frac{1}{\phi\lambda_{p}(1-\omega)}\bigg(1-(1+\omega\phi\lambda_{p})^{1-1/\omega}\bigg)
\label{eqn17}
\end{eqnarray}
are the modified radiosensitive parameters of the model.

\subsection{Model Fitting  Parameters}

To obtain the over-dispersion parameter that maximizes the
likelihood of making the observations given the parameters, we fit the NB distribution (Eq. (\ref{eqn5}) with $\omega =r$) to the survival data on 1.15 MeV proton
irradiated V79 cells and 25 MeV helium ion irradiated T1 cells
\cite{Prise1990}. A log-likelihood maximization was performed
across the analysed cells to find the best-fit value for a
particular data set. The NB log-likelihood function was performed
using the parmhat function that specifies control parameters for
the iterative algorithm the function uses. The programming was
implemented in MATLAB $\copyright$R2019a software. At $95\%$
confidence intervals, the value of $r$ was found to be 0.043 and
0.241, for the lower and upper limits, respectively. The bare overdispersion parameter was set to a constant value of 0.142, the best-fit
value for data on V79 cell survival. \\
\begin{figure}[!ht]
\begin{center}
\begin{tabular}{cccccc}
\scalebox{0.60}{\includegraphics{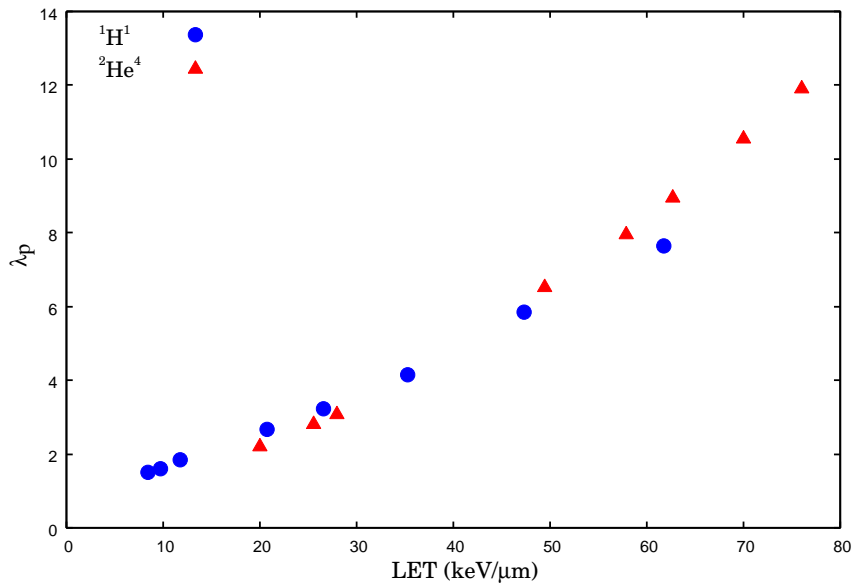}}
\end{tabular}
\caption{\label{fig1} DSBs yielded by each primary particle that cause DSB, $\lambda_{p}$, of V79
cells irradiated with protons and helium ions.}
\end{center}
\end{figure}

\noindent The yield of DSBs induced by hadrons and light ions was directly calculated
with fast Monte Carlo damage simulation (MCDS) software
\cite{MCDScode}. This algorithm captures the trend in the DNA damage
spectrum with the possibility that the small-scale spatial
distribution of elementary damages is governed by stochastic
events and processes \cite{Stewart2015,Stewart2018}. It has been
observed that the MCDS algorithm gives reliable results of the damage
yields that are comparable to those obtained from computationally
expensive but more detailed track structure simulations. For MCDS
simulations, the results of DNA damage yields for protons and
light ions are usually obtained within minutes. The random Gaussian error is added by repeating the
simulation 10 times for each LET value to generate ten
ensembles of DSB yield measurements for each LET value for later
analysis. The expectation values and statistical errors were estimated using the jackknife method. With $Y$ and $\lambda$ obtained with MCDS, and $\omega$ obtained as above, $n_{p}$ and $\lambda_{p}$ were obtained by using Eqs. (\ref{eqn8a}) and (\ref{eqn8}). \\
\noindent Fig.\ref{fig1} shows the average number of DSBs yielded by each primary particle that
cause DSB, $\lambda_{p}$, as a function of LET. For a constant dose, DSB yield per track changes more quickly with LET. This is as expected since the number of primary particles passing through the cell nucleus is inversely proportional to LET. The results show that for protons at LET $\leq$ 10 keV/$\mu$m, $\lambda_{p}$ increases rather slowly and has similar behaviour to that of photons in this LET region. This may be one of the reasons why the relative biological effectiveness of low LET protons is similar to that of photons. With protons, a two-fold increase in $\lambda_{p}$ is observed when LET was increased from 10 to 60 keV/$\mu$m. whereas $\lambda_{p}$ increases relatively quickly with helium ions at high LET, the difference between $\lambda_{p}$ values for protons and helium ions is negligible for $ 20 <$ LET$<30$ keV/$\mu$m. This is expected since ion LET is understood to modulate radiosensitivity through different patterns of energy deposition: higher LET radiations are more densely ionizing, which results in an increased DNA-double strand break (DSB) yield and
clustered DSB yield per unit dose \cite{Friedland2017}. Also, for the same LET, different ions will have differing DNA damage yields due to differing track ionization densities, e.g., a He-ion will have a higher DSB and clustered DSB yield compared to a C-ion of the same LET due to its much denser track \cite{Friedland2017}. \\
\begin{table}[ht!]
\begin{center}
\caption{Best fit parameters of the model for V79 cell }
\label{tab1}
\begin{tabular}{ccccccc}\hline\hline
Parameters & NB model & NB($\chi^{2}/N_{df}$) & Poisson model & Pois($\chi^{2}/N_{df}$) \\ \hline
Cell radius & $5.4 \pm 0.2$ $\mu$m & & & \\
Overdispersion $r$ & 0.146 & & & \\
$\mu_{x}$ & $0.9561 \pm 0.0235$ &0.872 & $0.9794 \pm 0.0163$ & 0.931\\
$\phi$ & $0.0608\pm 0.0381$ & & $0.0593\pm 0.0212$ & \\
$\xi$ & $0.0398 \pm 0.0129$ & & $0.0412\pm 0.0209$ & \\
$\eta\lambda_{p}\rightarrow 1$ & $(9.386\pm 0.104)\times 10^{-4}$ & 0.764 & $(9.785\pm 0.120)\times 10^{-4}$ & 0.886\\
$\eta\lambda_{p}\rightarrow \infty$ & $ 0.0065\pm 0.0001$ & 0.812 & $ 0.0068\pm 0.0001$ & 0.823\\ \hline
\end{tabular}
\end{center}
\end{table}

\noindent The parameters $\mu_{x}, \phi$, and $\xi$ were obtained by fitting the experimental data of cell survival curves of V79 cells \cite{Furusawa2000}. Twenty data sets for cell survival were selected to cover a wide range of high, medium, and low surviving fractions, allowing a detailed analysis of survival curve shapes. GRABIT Data Figure Digitizer routine in MATLAB was used to estimate data points and error bars from published graphs. First, the parameters $\mu_{x}$, $\phi$ and $\xi$ were obtained by fitting the experimental data of $\alpha$ values and the calculated $n_{p}$ and $\lambda_{p}$ values with Eq.(\ref{eqn17a}). Using these values together with the experimental data of cell survival for V79 cells exposed in $X$-rays, we obtained $\eta_{\lambda_{p}\rightarrow 1}$ with Eq.(\ref{eqn16}). Finally, $\eta_{\lambda_{p}\rightarrow \infty}$ was obtained by fitting the experimental data with Eq.(\ref{eqn17}). The goodness of the fits was gauged by the reduced chi-squared test using the GNU plot. The best fit values and their uncertainties are summarised in Table \ref{tab1}.

\section{Results and Discussion}
\begin{figure}[!ht]
\begin{center}
\begin{tabular}{ccccccccccc}
\scalebox{0.45} {\includegraphics{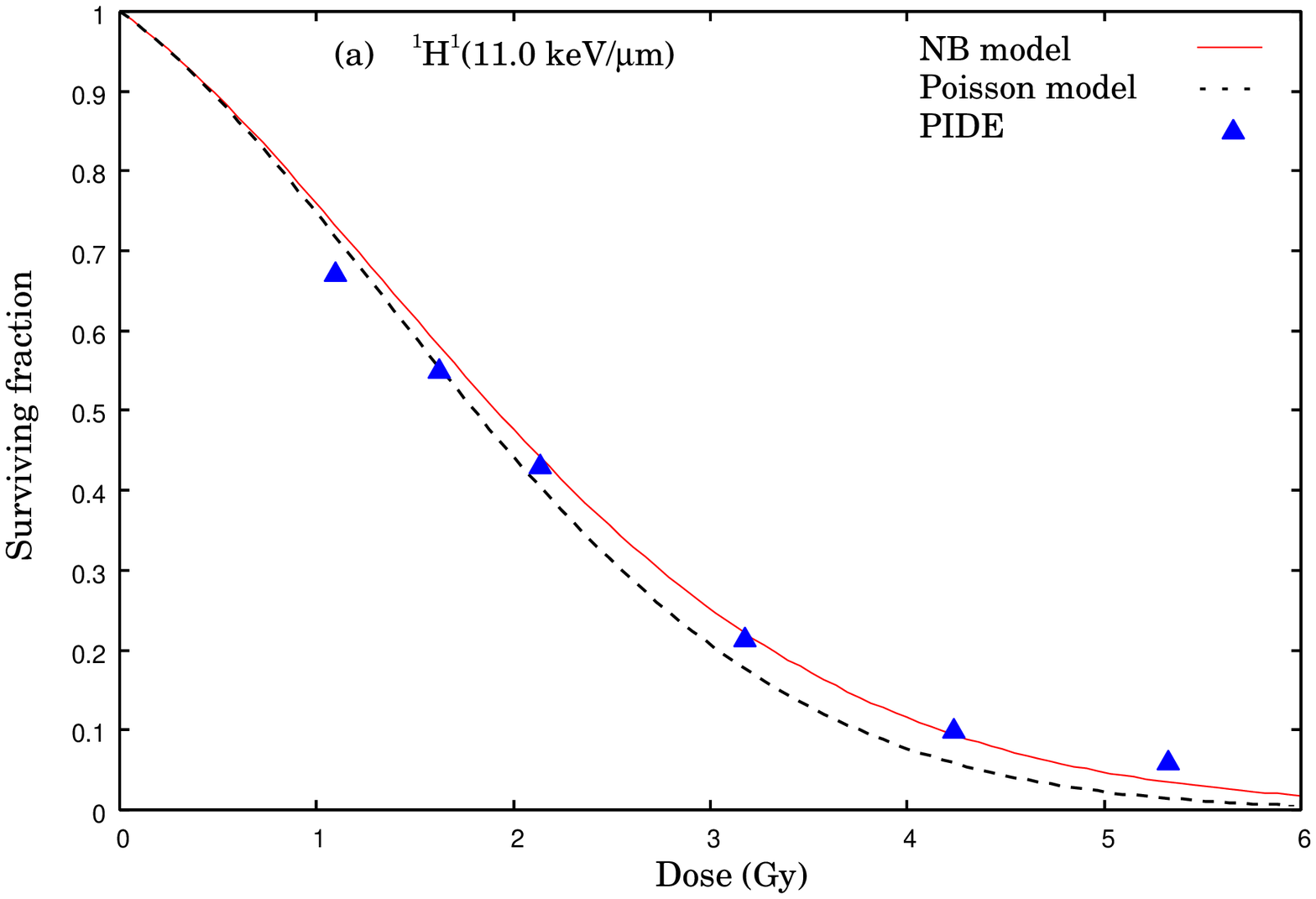}} &
\scalebox{0.45}{\includegraphics{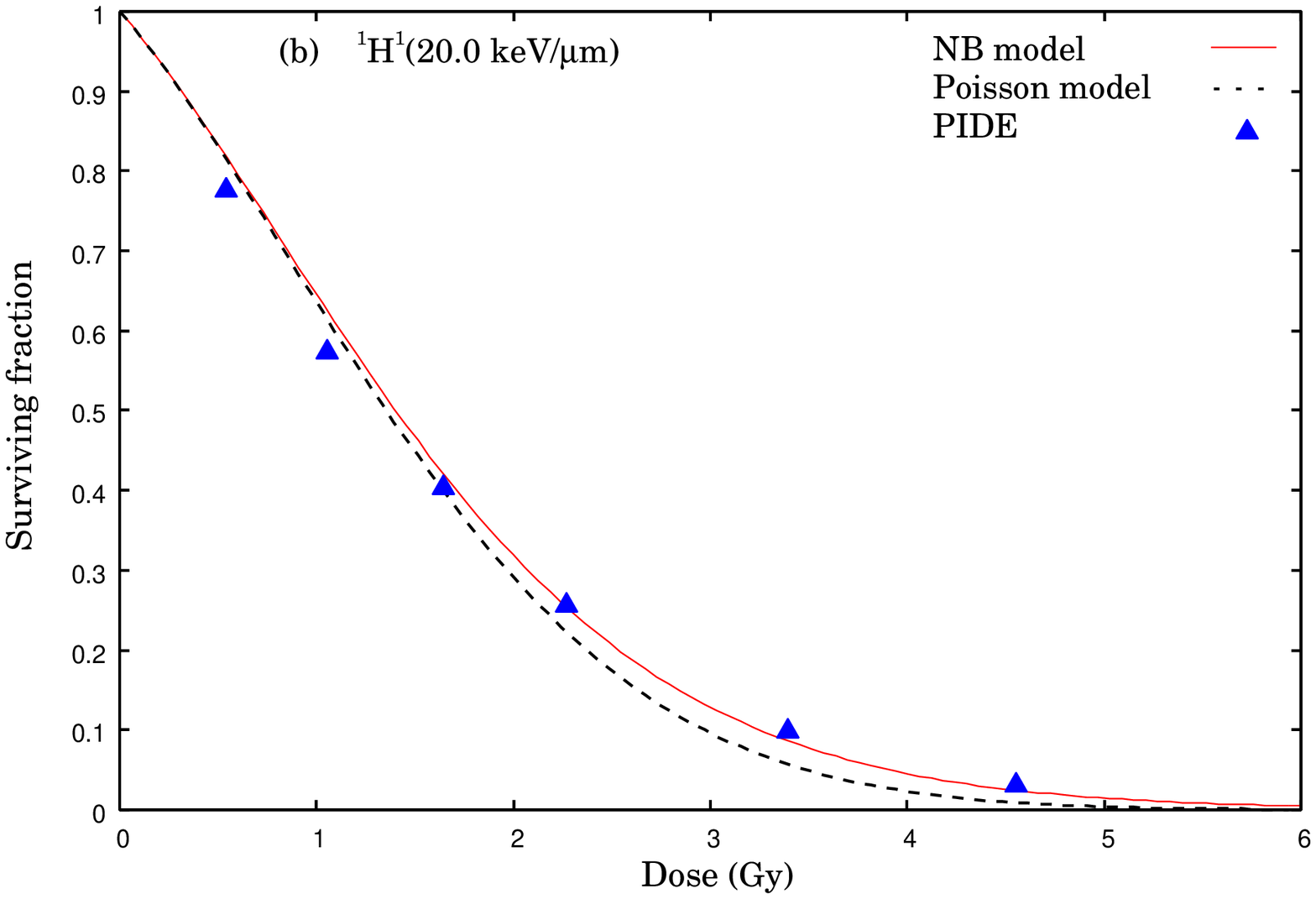}}\\
\scalebox{0.45} {\includegraphics{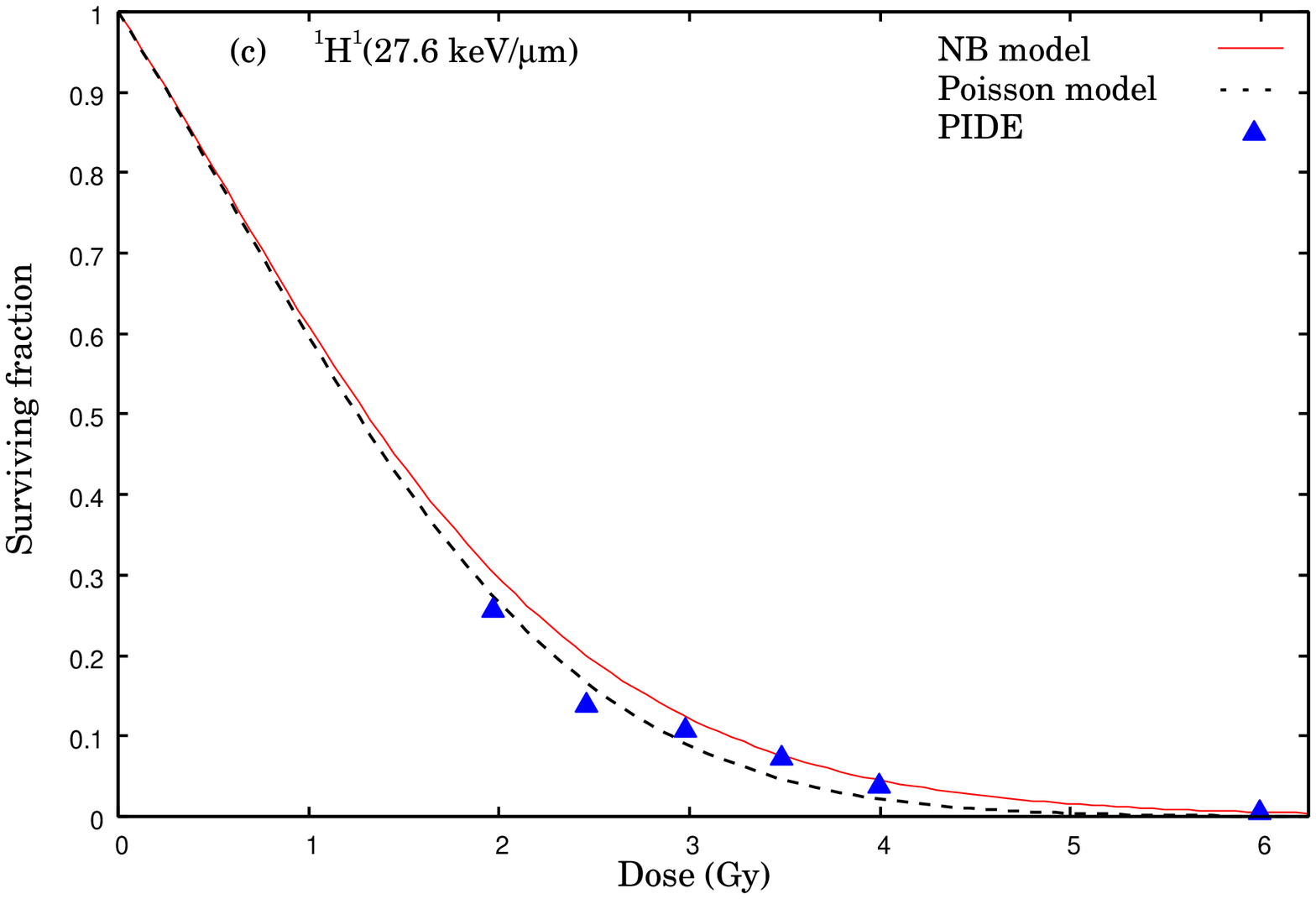}} &
\scalebox{0.45}{\includegraphics{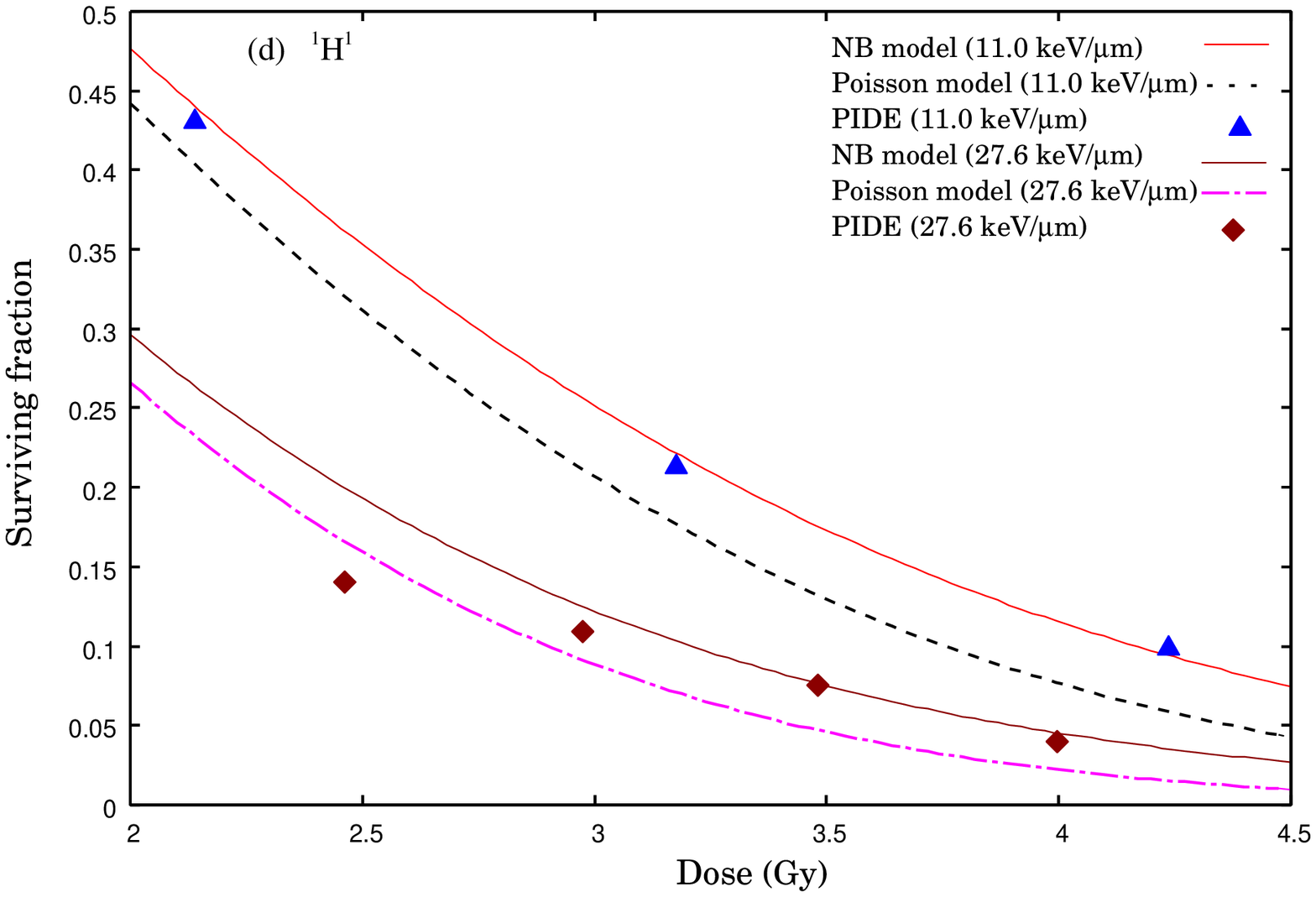}}
\end{tabular}
\caption{\label{fig2} Comparison between measured (PIDE \cite{PIDE2013} - solid symbols) and model cell survival fractions (solid and dashed lines) for V79 cells irradiated by protons at different LET. Magnified dose-response curves for improved visualization of the modelled fits are shown in panel d.}
\end{center}
\end{figure}

Using the parameters obtained above, we evaluate and display the effect of overdispersed lesions for V79 cells, irradiated by protons at low and high doses for different LET values, in Fig.\ref{fig2}. Comparing the NB and Poisson model dose-response predictions, (Fig.\ref{fig2} a-c), to the data in the Particle Irradiation Data Ensemble (PIDE) \cite{PIDE2013}, we observe that the Poisson model agrees well with the experimental data at doses $\leq$ 2 Gy but underestimates the “lower tail” of the observed distribution at medium and higher doses. On the other hand, NB error distribution outperforms the Poisson model and shows a good agreement with the experimental values at medium and higher doses for all LET values explored here.
\noindent A closeup of the dose-response curve for improved visualization of the modelled fits (Fig.\ref{fig2}-d) shows straightening of dose-response in the intermediate dose region. This indicates that overdispersion causes the dose-response curves to approximate log-linear behaviour at high doses. This is because overdispersion alters the relationship between the mean number of lethal lesions per cell and survival (the probability of a cell having zero lethal lesions). It is obvious that with the same mean number of lethal lesions per cell, NB distribution predicts a larger probability of a cell to contain zero lethal lesions than that predicted by the Poisson distribution. This effect is expected to increase with the increasing mean number of lethal lesions per cell and with increasing overdispersion as the damage saturation correction implies that a further increase in damage within tracks does not further increase cell killing. The important consequence is that at radiation doses that kill most cells by producing a large mean number of lethal lesions per cell, the Poisson distribution can predict substantially lower cell survival probabilities than the overdispersed NB distribution.\\

\begin{figure}[!ht]
\begin{center}
\begin{tabular}{cccccccc}
\scalebox{0.45} {\includegraphics{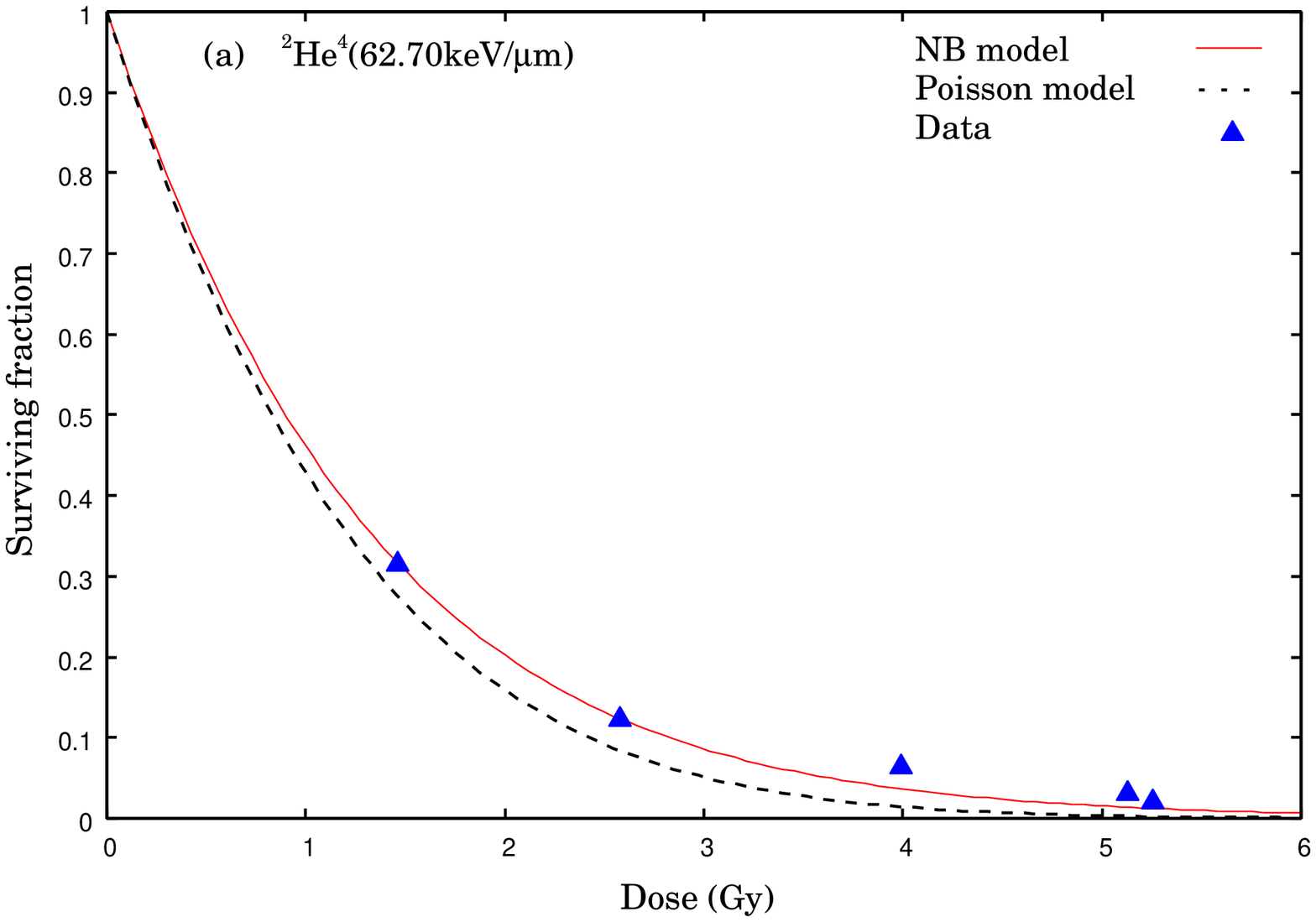}} &
\scalebox{0.45} {\includegraphics{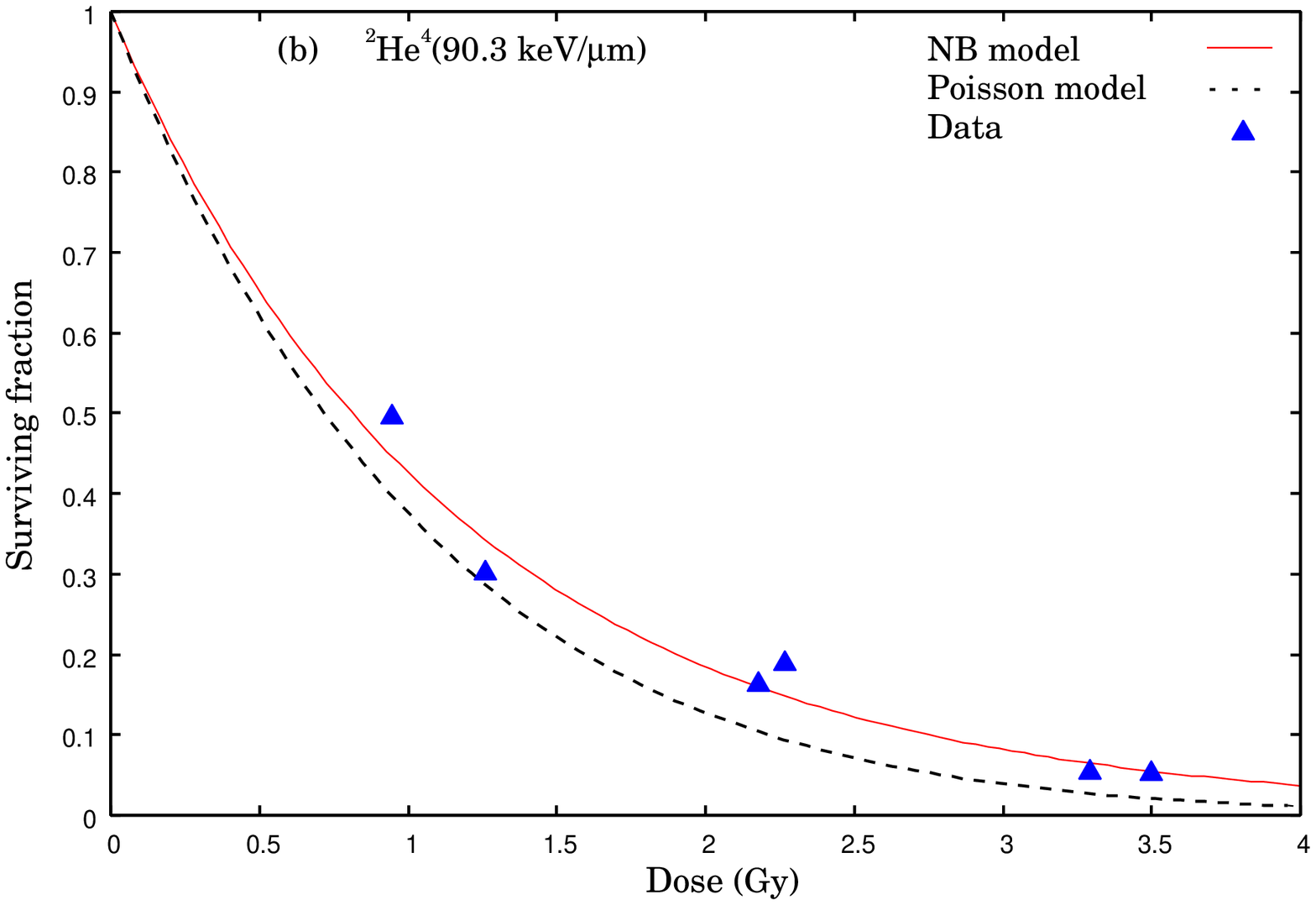}} \\
\scalebox{0.45} {\includegraphics{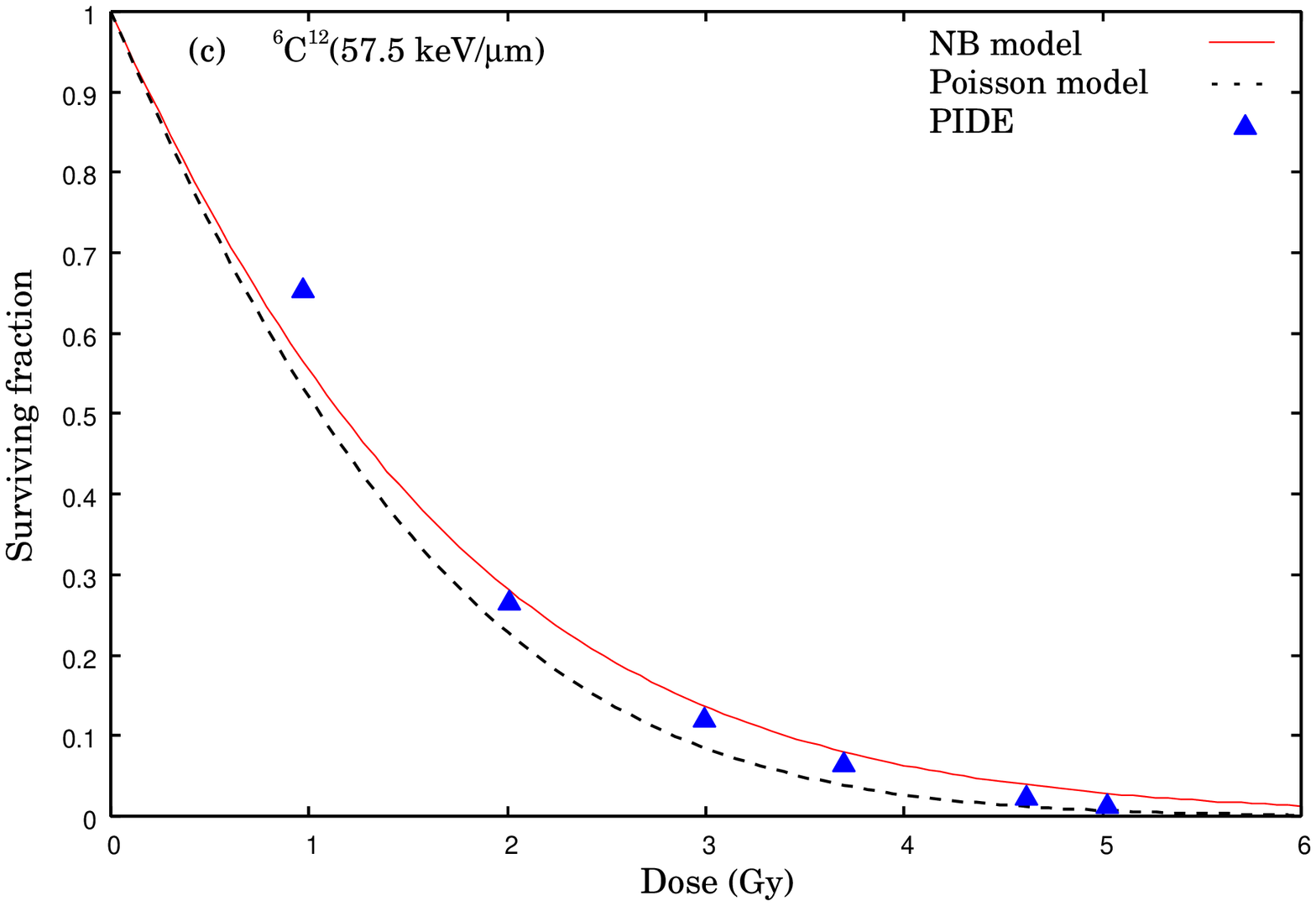}} &
\scalebox{0.45} {\includegraphics{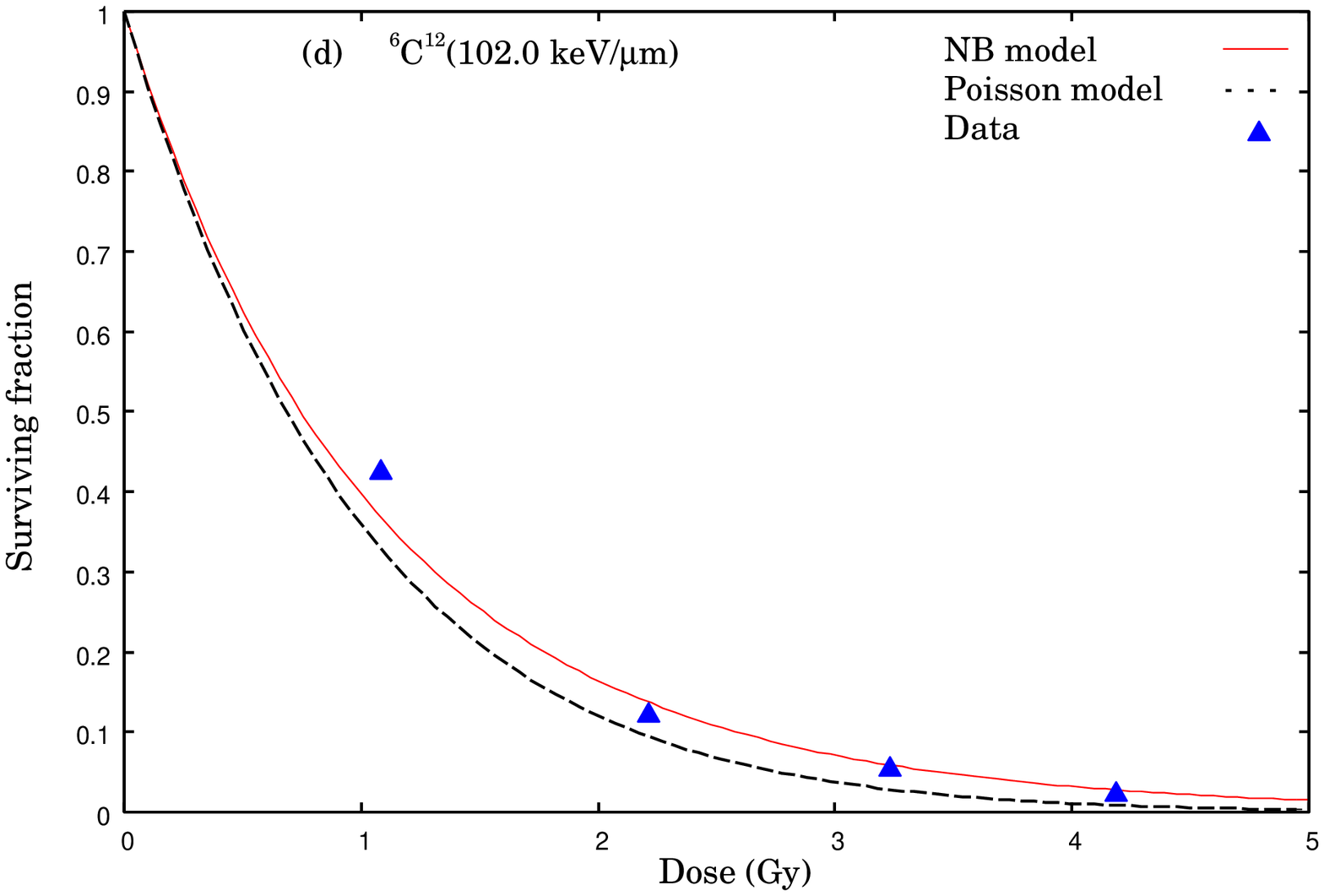}}
\end{tabular}
\caption{\label{fig3} Comparison between experimental data (PIDE \cite{PIDE2013}, Data \cite{Tilly1999}) of cell surviving fractions (points) and modelled results (solid and dashed lines) for V79 cells irradiated by helium and carbon ions.}
\end{center}
\end{figure}

\noindent For the same cell type, the NB error model showed enough flexibility to describe the surviving fraction not only for the particle types used in model parameter fitting but also for different types of particles at different LET. Fig.\ref{fig3} shows the evidence for overdispersion of lethal lesions was also found at relevant doses for V79 cells irradiated by helium and carbon ions at different LET.
The effective plots (Fig.\ref{fig3} - a and b) show that the NB error distribution predictions with helium ions are considerably closer to the measured cell survival data \cite{PIDE2013,Tilly1999} at doses $>$ 2 Gy. To quantify the agreement between the predicted and measured survival fractions, we evaluated the Poisson and NB error modelled cell survival probabilities at various doses. In contrast to NB distribution, the Poisson modelled survival curves predict 27.8, 8.5, 1.37, and 0.3$\%$ survival fractions at 1.47, 2.58, 4.0, and 5.25 Gy relative to 31.9, 12.4, 6.2, and 2.06$\%$ measured fractions, respectively. The corresponding zero lethal lesion probabilities predicted by the NB error model are 31.9, 12.4, 3.67, and 1.37$\%$ are considerably closer to the measured data. The comparison reveals that the Poisson model tends to produce fits away from the measured values at higher doses of helium ions. These results provide support for our contention that the distribution of lethal chromosomal lesions per cell at high doses of ionizing radiation may not be optimally described by the Poisson distribution. Alternative approaches such as the NB distribution, which allows the variance to be larger than the mean, allow better fits to such data. The added flexibility can translate into better descriptions of available survival dose-response data, and better predictions of the dose responses at low doses/ dose rates, as well as at high doses. For example, moderate overdispersion of lethal lesions per cell modelled by the NB distribution substantially modifies BED estimates for radiotherapy regimens that use three to five dose fractions of $\geq$ 5 Gy/fraction. \\

\noindent Our analysis of dose-response with carbon ions at 57.7 and 102 keV/$\mu$m shows that Poisson modelled cell survival trends moderately underestimate the actual survival data at medium doses. This is in contrast to the cell survival probabilities with proton and helium ions, the Poisson distribution describes the measured data for the doses $\geq$ 4.65 Gy. This might be due to the reason that the ionization density of carbon ions induces more complex DNA damage which makes the DNA repair process complex, partially due to retarded enzymatic activities, leading to increased chromosome aberrations and cell
death. This suggests that, in general, the repair process following heavy-ion exposure is LET-dependent, but with nonhomologous end-joining
defective cells, this trend is less emphasized.

\section{Conclusions}

We adopted the effect of a customized negative binomial distribution as a more flexible approach that accounts for
the non-random clustering and overdispersion of lethal lesions and evaluated the modifications in shapes of cell survival response curves.
The NB error distribution response model was tested by fitting the cell survival predictions to the measured data for V79 cells irradiated with LETs within the range of 11– 102 keV/$\mu$m. The model provides good agreement with experimental cell survival data in the medium and high dose regions for the range of LET investigated, confirming the effectiveness of the parameterization method. These results from lymphocytes and fibroblasts provide support for our contention that the distribution of lethal chromosomal lesions per cell at high doses of  radiation  may not be optimally described by the Poisson distribution. Alternative approaches, such as the NB distribution, which allows the variance to be larger than the mean, allow for better fitting of such data. Although there are other explanations for cell survival curve straightening at high doses, our results do not contradict these explanations, since a non-Poisson error distribution of lethal lesions and a non-LQ dose dependency are not mutually exclusive. Therefore, we do not choose to advocate one over the other. We argue, however, that even with a continuously curving dose-response formalism and an overdispersed error distribution straightening of survival curves at high doses can be explained without causing other fundamental questions about the relationship between dose and average number of lethal events. Taking the error distribution into consideration may be a useful approach for modelling the effects of modern radiotherapy with high doses per fraction. Such adjustments within an improved NB distribution can potentially improve radiobiological modelling at high-dose cancer radiotherapy and may be useful for optimization of radiotherapy treatment planning to maximize tumour control.

\section*{Conflict of Interest}
No conflict of interest to declare.

\appendix
\section{}

We use a customized negative binomial (NB) distribution for
evaluating the probability of observing $k$ lethal lesions in a
cell
\begin{equation}
P_{NB}(k) = \frac{\Gamma (k+1/\omega )}{\Gamma (1/\omega)\times k!}
\bigg(\frac{1}{1+\omega\lambda }\bigg)^{1/\omega} \times
\bigg(\frac{1}{1+1/\omega\lambda}\bigg)^{k}.\nonumber
\end{equation}
The likelihood function for $N$ iid observations $(k_{1}, k_{2},
\cdots k_{N})$ is given by
\begin{displaymath}
L_{NB}=\prod_{i=1}^{N}f(k_{i})
\end{displaymath}
from which the log-likelihood function can be written as:
\begin{eqnarray}
l_{NB} & =& - \bigg(\frac{1}{\omega}\bigg)\times (1+k+\omega)\times
\ln(1+\omega+\lambda)+ \nonumber\\
& & k\times \bigg(\ln(\omega)+\ln(\lambda )\bigg)+ \ln
\left[\Gamma\bigg(\frac{1+k+\omega}{\omega}\bigg)\right]- \ln[\Gamma (1+k)]
- \ln\left[\Gamma\bigg(\frac{1}{\omega}\bigg)\right].\nonumber
\end{eqnarray}

Assuming that the correct repair distribution is NB distributed,
the probability that a break will be repaired correctly is
\begin{eqnarray}
P_{NB} & =&
\sum_{k=0}^{\infty}\frac{1}{1+k}P_{NB}(k)=\sum_{k=0}^{\infty}
\frac{\Gamma (k+\omega )}{\Gamma (\omega
)} \times \frac{1}{1+k} p^{\omega}(1-p)^{k}\nonumber\\
 & = & \sum_{k=0}^{\infty}
 \frac{(k+\omega-1)!}{(\omega -1)!\,(k+1)!}p^{\omega}(1-p)^{k}. \nonumber
\end{eqnarray}
If we let $k+1=s$ and $\omega-1=z$, then
\begin{eqnarray}
P_{NB}& = & \frac{p}{z(1-p)}\sum_{s=1}^{\infty}\frac{(s+z-1)!}{s!\,(z-1)!}p^{z}(1-p)^{s}\nonumber\\
& = & \frac{p\omega}{(1-\omega )(1-p)}(1-p^{\frac{1}{\omega} -1}).
\label{Aeqn1}
\end{eqnarray}
By fixing the average number of breaks equal to the expected
interaction rate $\lambda (= \eta (\lambda_{p})n_{p})$, we can use
this to calculate the total probability of misrepair as a function
of $\lambda_{p}$. Using $p= 1/(1+\omega\lambda )$, we expand Eq.
(\ref{Aeqn1}) to obtained
\begin{equation}
P_{NB} = \frac{1}{\lambda
(1- \omega )}\left[1-\bigg(1+\omega \lambda\bigg)^{1-1/\omega}\right].\nonumber
\end{equation}

\end{document}